\documentclass[reprint,aps,pre,twocolumn,amsmath,amssymb, superscriptaddress]{revtex4-1}
\usepackage[english]{babel}
\usepackage{graphicx}
\usepackage{epstopdf}
\usepackage{amsmath}
\usepackage[rightcaption]{sidecap}
\usepackage{dcolumn}
\usepackage{bm}
\usepackage{color}
\usepackage{sidecap}
\usepackage{booktabs}

\begin{document}

\title{Multiscale permutation entropy analysis of laser beam wandering in isotropic turbulence}

\author{Felipe Olivares}
\email{olivaresfe@gmail.com}
\affiliation{Instituto de F\'isica, Pontificia Universidad Cat{\'o}lica de Valparaiso (PUCV), 23-40025, Valpara\'iso, Chile}
\author{Luciano Zunino}
\email{lucianoz@ciop.unlp.edu.ar}
\affiliation{Centro de Investigaciones \'Opticas (CONICET La Plata - CIC), C.C. 3, 1897 Gonnet, Argentina,}
\affiliation{Departamento de Ciencias B\'asicas, Facultad de Ingenier\'ia, Universidad Nacional de La Plata (UNLP), 1900 La Plata, Argentina}
\author{Dami\'an Gulich}
\email{dgulich@iflysib.unlp.edu.ar}
\affiliation{Departamento de Ciencias B\'asicas, Facultad de Ingenier\'ia, Universidad Nacional de La Plata (UNLP), 1900 La Plata, Argentina}
\affiliation{Instituto de F\'isica de L\'iquidos y Sistemas Biol\'ogicos, CONICET, UNLP, Calle 59 Nro. 789, 1900 La Plata, Argentina}
\author{Dar\'io G. P\'erez}
\email{dario.perez@pucv.cl}
\affiliation{Instituto de F\'isica, Pontificia Universidad Cat{\'o}lica de Valparaiso (PUCV), 23-40025, Valpara\'iso, Chile}
\author{Osvaldo A. Rosso}
\email{oarosso@gmail.com}
\affiliation{Instituto de F\'isica, Universidade Federal de Alagoas (UFAL), BR 104 Norte km 97, 57072-970. Macei\'o, Alagoas, Brazil}
\affiliation{Instituto Tecnol\'ogico de Buenos Aires (ITBA), C1106ACD, Av. Eduardo Madero 399. Ciudad Aut\'onoma de Buenos Aires, Argentina}
\affiliation{Complex Systems Group, Facultad de Ingenier\'ia y Ciencias Aplicadas, Universidad de los Andes, Av. Mons. \'Alvaro del Portillo 12.455, Las Condes, Santiago, Chile}

\date{\today}        

\begin{abstract}
We have experimentally quantified the temporal structural diversity from the coordinate fluctuations of a laser beam propagating through isotropic optical turbulence. The main focus here is on the characterization of the long-range correlations in the wandering of a thin Gaussian laser beam over a screen after propagating through a turbulent medium. To fulfill this goal, a laboratory-controlled experiment was conducted in which coordinate fluctuations of the laser beam were recorded at a sufficiently high sampling rate for a wide range of turbulent conditions. Horizontal and vertical displacements of the laser beam centroid were subsequently analyzed by implementing the symbolic technique based on ordinal patterns to estimate the well-known permutation entropy. We show that the permutation entropy estimations at multiple time scales evidence an interplay between different dynamical behaviors. More specifically, a crossover between two different scaling regimes is observed. We confirm a transition from an integrated stochastic process contaminated with electronic noise to a fractional Brownian motion with a Hurst exponent $H=5/6$ as the sampling time increases. Besides, we are able to quantify, from the estimated entropy, the amount of electronic noise as a function of the turbulence strength. We have also demonstrated that these experimental observations are in very good agreement with numerical simulations of noisy fractional Brownian motions with a well-defined crossover between two different scaling regimes.

\end{abstract}
%

\maketitle

\section{Introduction}

The random nature of the turbulent atmosphere affects the propagation of any light beam propagating through it. Indeed, the turbulent velocity field shuffles and breaks pockets of air with different index of refraction; in doing so, the index fluctuations inherits the stochastic properties of the turbulent velocity field. Consequently, any wavefront emerging from a turbulent region will experiment random distortions. Schwartz {\it{et al.}} \cite{Schwartz1994} hypothesized that turbulence-degraded wavefront phase fluctuations can be modeled by a fractional Brownian motion (fBm) with a Hurst exponent $H = 5/6$, within the Kolmogorov model. Later was proven \cite{Perez2004B,Perez2008} that wavefront phase can be correctly described by (2D) Levy fBm field in the inertial range. Fractional Brownian motion is a family of Gaussian self-similar stochastic processes with stationary increments. The former is a well accepted model for fractal phenomena that have an empirical spectra of power-law type $1/f^{\alpha}$ and $\alpha = 2H+1$ with $1<\alpha <3$ \cite{Mandelbrot1968}. Its long-range correlations are quantified by the Hurst exponent $H$ $\in (0, 1)$. These processes exhibit temporal memory for any value of $H$ except for $H = 1/2$, which corresponds to classical Brownian motion (random walk). Thus, the Hurst parameter defines two distinct regions in the interval $(0, 1)$. When $H > 1/2$, consecutive increments tend to have the same sign so that these processes are persistent. For $H < 1/2$, on the other hand, consecutive increments are more likely to have opposite signs, and the underlying temporal dynamics are anti-persistent \cite{Feder1988}.

Any laser beam that propagates through the turbulence experiments perpendicular displacements to the initial unperturbed direction of propagation \cite{Andrews}. These displacements emerge from the beam phase fluctuations. This phenomenon is commonly known as {\textit{laser beam wandering}} because of the dancing that the beam performs over a screen. Since it is very sensitive to the turbulence behavior, it has been used in different experimental configurations to estimate the characteristic scales and parameters associated with the turbulence, such as the inner scale $l_0$, the outer scale $L_0$ and the index of refraction structure constant $C_n^2$ \cite{Andrews,Zhang2001,Innocenti2005,Funes2007}. Recently, Zunino {\it{et al.}} \cite{Zunino2014} experimentally confirmed the $5/6$-exponent for the angle-of-arrival fluctuations of stellar wavefronts propagating through atmospheric turbulence. Later on, they \cite{Zunino2015B} experimentally showed that the Hurst exponent can be estimated from the coordinate fluctuations of the laser beam propagating through fully developed isotropic turbulence without assuming a particular spectral behavior; independent of any theoretical model, proving to be $H=5/6$.

Usually, the information captured from a laser beam wandering experiment are temporal records of the coordinates and intensity fluctuations. From these measurements or commonly called time series (TS), is possible to characterize the underlying turbulence dynamics \cite{Zunino2015B,Funes2016}. Numerous methodologies focus on the estimation of entropic quantifiers to characterize the dynamical behavior of a given system from a TS. In this scenario a probability distribution function (PDF) is assumed \textit{a priori}. However, the implementation of an appropriate methodology, \textit{i.e.} one that extracts all the relevant intrinsic dynamical information, for estimating the PDF related to a TS is a subtle issue. Actually, it depends on particular characteristics of the data, such as stationarity, length of the TS, level of noise contamination, etc. Many schemes have been proposed for a proper estimation of the PDF associated with a TS; without being exhaustive, we can mention: binary symbolic dynamics \cite{Mischaikow1999}, Fourier analysis \cite{Powell1979}, wavelet transform \cite{Rosso2002}, permutation analysis \cite{Bandt2002}, Lempel-Ziv permutation analysis \cite{Zozor2013}, etc. In particular, the permutation analysis maps a raw TS into a corresponding sequence of symbols called ordinal patterns \cite{Bandt2002}. This symbolic method is simple, robust, and, the most important fact is that it takes notice of time causality in dealing with the dynamics of the system. No model assumptions are needed. Distinction between deterministic (chaos) from stochastic nature of TS, as well as, identification of dynamic changes at different temporal scales have been accomplished by this symbolic approach \cite{Rosso2007, Olivares2012,Zunino2012, Zunino2015,Olivares2016}. In particular, ordinal patterns and permutation entropy have become a significant advance in the characterization of fBm and its increments \cite{Rosso2007B, Zunino2008, Olivares2016}. 

In this paper, we applied the ordinal patterns analysis to estimate the entropy of the fluctuations of a laser beam centroid propagating through an isotropic optical turbulent medium. Here, we aim to go beyond from a previous work \cite{Zunino2015B}, by implementing a multiscale analysis of the permutation entropy that allows us to characterize all the dynamical information contained in the temporal measurements. We confirm the presence of an interplay between two different dynamical behaviors: an integrated stochastic process for small time scales and a fBm for large temporal scales. As the turbulence intensity increases, the permutation entropy estimated from the experimental records, converges to the value associated with the permutation entropy of a fBm with a Hurst exponent $H=5/6$, as expected.  In addition, we are able to quantify the amount of noise contamination in the measurements.

The remainder of this paper is structured as follows: the experimental system is introduced in Section \ref{EE}. In Section \ref{PE},  we briefly describe the ordinal patterns and permutation entropy. Experimental results and numerical simulations are presented and discussed in Section  \ref{RR}. Finally, some concluding remarks are given in Section \ref{CC}.

\section{Experiments} \label{EE}

\begin{figure} [!ht] 
\centering
\includegraphics{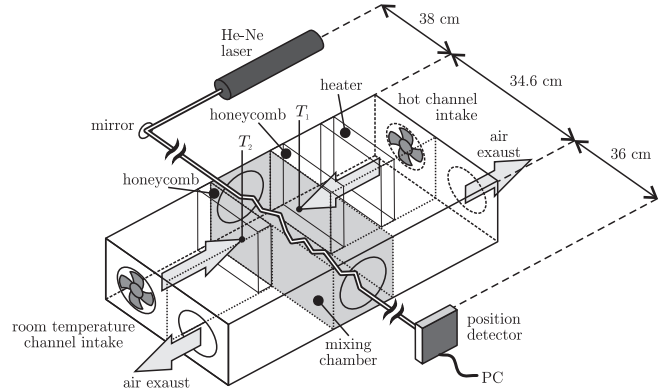}
\caption{Schematic diagram of the laboratory experimental setup.} \label{fig1}
\end{figure}
Controlled turbulent air flow is generated using a device commonly called turbulator. Similar to the originally proposed by Fuchs {\it{et al.}} \cite{Fuchs1996}, and later enhanced by Keskin {\it{et al.}} \cite{Keskin2006}. Briefly, the optical turbulence extending over a 35 cm channel is the product of the collision of two masses of air, one hot and one cold, that are pushed by identical fans through honeycombs placed opposite to each other at the sides of this chamber. Each fan is spinning at equal velocities so the turbulence characteristics are only due to the temperature difference between the two masses of air. The hot source has an electric heater controlled by changing the current passing through it. By increasing the temperature of the hot source, different turbulent intensities can be produced \cite{Jolissaint2001,Gulich2016}. The turbulator is a fully characterized single turbulent layer that offers repeatability. The strength of the artificial turbulence is quantified through the index of refraction structure constant $C_n^2$. In this characterization, $C_n^2$  is expressed as a function of the temperature difference between hot and cold sources ($T_1$ and $T_2$, respectively, in Fig. \ref{fig1}) \cite{Masciadri1997}. The inner and outer scales are not related to the strength of the turbulence.

The experiment was performed in controlled conditions in which a laser beam propagates through the artificial turbulence---see Fig. 1 for a schematic view of the optical setup. The wandering of the laser beam (10 mW HeNe Melles Griot Model 05-LHP-991) is detected by a position-sensitive detector with an area of 1 cm$^2$ (UDT SC-10 D). This detector measures the position of the centroid of the impinging laser beam with an accuracy of 2.5 $\mu$m, so very small position deflections can be detected. Fluctuations on the centroid position along the vertical and horizontal axes were recorded at 2 kHz. Experiments with 13 temperature differences $\Delta T = T_1 -T_2$ ranging from 5 $^{\circ} C$ to 180 $^{\circ} C$ were performed. For further details about the experiment see Ref. \cite{Zunino2015B}.

\section{Permutation entropy} \label{PE}

Bandt and Pompe (BP) introduced a symbolic methodology, which arises naturally from a given time series without any model assumptions \cite{Bandt2002}. ``Partitions'' of length $D$ are devised by appropriately ranking the neighboring series' values rather than allocate amplitudes according to different levels. Given a one-dimensional TS, ${\mathcal{X}} = \{x_t ; t = 1, \dots,M\}$ first one has to choose two parameters, the embedding dimension $D \geqslant 2$ ($D \in {\Bbb N}$, the pattern length) and the embedding delay $\tau$ ($\tau \in {\Bbb N}$, the time separation between the values).  After that, the time series is partitioned into subsets of length $D$ of consecutive ($\tau = 1$) or non-consecutive ($\tau > 1$) values, generated by $(t) = (x_t, x_{t+\tau}, ..., x_{t+(D-2)\tau}, x_{t+(D-1)\tau})$, which assigns to each time $t$ the $D$-dimensional vector of values at times $t$, $t + \tau , .... , t + (D-1)\tau$. Clearly, more temporal information is incorporated into the vectors as the $D$-value increases. Then, each element of the vector is replaced by a number from zero to $D-1$ related to their original temporal position in the partition. By ordinal pattern related to the time ($t$) we mean the permutation $\pi_{i} = (r_0, r_1, ... , r_{D-1})$ of $[0, 1, . . . , D - 1]$, in accordance with the relative strength of each element in the ordered vector from low to high. Equal values in the TS are usually ranked according to their temporal order. This is justified if the values of ${\mathcal{X}}$ have a continuous distribution so that equal values are very unusual.

The pattern length, $D$, plays an important role in the evaluation of the appropriate PDF because it determines the number of accessible states or ordinal patterns, $D!$, and also conditions the minimum acceptable length $M \gg D!$ of the TS that one needs in order to work with a reliable statistics \cite{Bandt2002}. By counting the number of times each possible permutation $\pi_{i}$ appears in the symbolic sequence divided by the total number of vectors, one can compute the PDF of the ordinal patterns. The permutation entropy (PE) is just the classical Shannon entropy estimated by using this ordinal pattern probability distribution. Its normalized version is given by
\begin{equation}
	\mathcal{H}_D^{\tau} =  - \frac{1}{\ln(D!)} \sum_{i=1}^{D!} p(\pi_i) \ln(p(\pi_{i})).
\end{equation}

To illustrate the BP-recipe we will consider a simple example starting with a TS with seven ($M=7$) values $ {\mathcal{X}}  = \{4, 7, 9, 10, 6, 11, 3\}$, embedding dimension $D=3$ and embedding delay $\tau=1$. The first two triplet, (4, 7, 9) and (7, 9, 10), are mapped to the pattern ${012}$ since the values are originally placed in ascending order. On the other hand, (9, 10, 6) and (6, 11, 3) correspond to the pattern ${201}$ since $x_{t+2} < x_t < x_{t+1}$, while (10, 6, 11) is mapped to the ordinal pattern ${102}$ since $x_{t+1} < x_t < x_{t+2}$. Then, the associated probabilities with the six ordinal patterns are: $p({012}) = p({201}) = 2/5$; $p({102}) = 1/5$; $p({021}) = p({120}) = p({210}) = 0$.  Consequently, for this example, $\mathcal{H}_{3}^{1} =\frac{1}{\ln 6}( - 2(2/5) \ln(2/5) - (1/5) \ln(1/5) ) \approx 0.59$.

With respect to the selection of the parameters, BP suggest in their cornerstone paper to work with $3  \leq D  \leq  7$ and a time lag $\tau = 1$ \cite{Bandt2002}. Nevertheless, other values of  $\tau$  might provide additional information. It has been shown that this parameter is strongly related, when it is relevant, to the intrinsic time scales of the system under analysis \cite{Zunino2010,Soriano2011A,Soriano2011B,Zunino2012,Zunino2015,Olivares2016}. By changing the value of the embedding delay $\tau$ different time scales are being considered since $\tau$ physically corresponds to multiples of the sampling time of the signal under analysis. In this work, we report our results with an embedding delay $D=5$, yet qualitatively similar results were found by using $D=3$, 4 and 6. On the other hand, the embedding delay $\tau$ is varied between 1 and 50 for testing the underlying dynamics at several time scales. 

\section{Results and discussion}  \label{RR}

\begin{figure} [!ht] 
\centering
\includegraphics[scale=0.4]{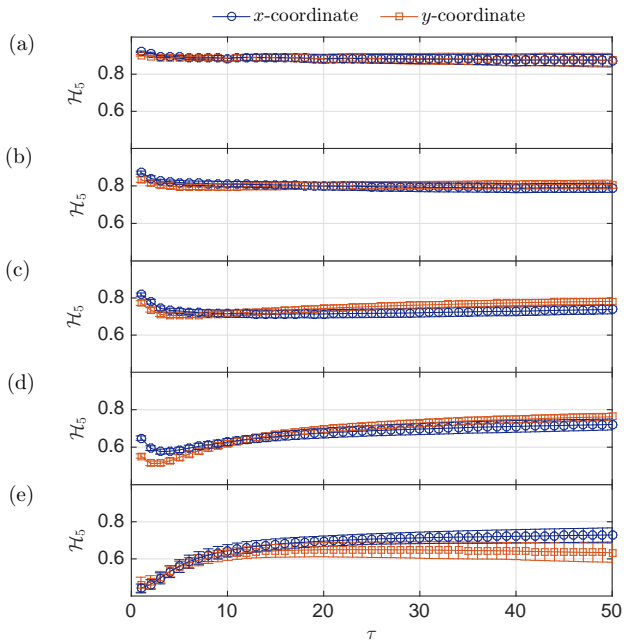}
\caption{Mean and SD (displayed as error bars) of the normalized PE  as a function of the embedding delay $\tau$, using an embedding dimension $D=5$, for the integrated fluctuations of the centroid coordinates of the laser beam for five representative turbulence strengths $C_n^2 [m^{-2/3} \times 10^{-9}]$ (a) 5.7, (b) $6.7$, (c) $8.0$,  (d) $14.1$, (e) $40.2$. Qualitative similar results are found with $D=3$, $4$ and $6$. } \label{fig2}
\end{figure}

Twenty-one independent realizations of $M=20,000$ points for each coordinate were recorded for each turbulent condition. The PE was estimated from the integrated sequences for both (horizontal and vertical) coordinates.  Since \textit{a priori} there is not a privileged time scale (\textit{i.e.} an optimal embedding delay $\tau$) at which the entropy should be estimated, we carried out a multiscale analysis. Figure \ref{fig2} shows mean and standard deviation (SD) of the normalized PE with $D=5$, for the twenty-one realizations, as a function of the embedding delay $\tau$. Five representative turbulent strengths are depicted for the sake of better visualization. Similar qualitative results are found with $D=3$, 4 and 6. This multiscale analysis confirms a transition of the PE to a stable value as the time scale $\tau$ increases, which seems to be more pronounced for stronger turbulent intensities, Fig. \ref{fig2}(e). Without having any prior information about the measurements, we can associate the entropy evolution observed at low time scales with either the omnipresence of noise (which is inherent to any temporal sequences of measurements), or the existence of a crossover between two scaling laws.

\begin{figure} [!ht] 
\centering
\includegraphics[scale=0.3]{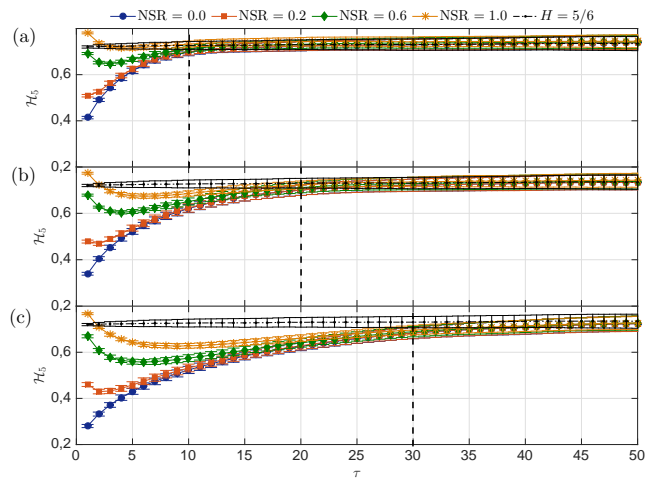}
\caption{Mean and SD (displayed as error bars) of the normalized PE  as a function of the embedding delay $\tau$, using an embedding dimension $D=5$, for a simulated fBm of length $M =20,000$ with a well-defined crossover at time scales (a) 10, (b) 20, and (c) 30. Different NSR are also included. Qualitative similar results are found when $D=3$, $4$ and $6$ are used.} \label{fig3}
\end{figure}

In order to better understand this issue, we have simulated twenty-one independent realizations of a fBm with $H=5/6$ using the MATLAB function {\it{wfbm}} (it uses the algorithm proposed by Abry and Sellan \cite{Abry1996}). The lengths of the simulated sequences are the same as the measured coordinate fluctuations ($M=20,000$). Furthermore, a well-defined crossover time scale $w$ was defined in the artificial TS, by dividing the series generated from the increments of the fBms (its derivative), into segments of length $w$. Then we integrate the segments, in such a way to obtain a process (of length $w$).  Thus, all the correlations for $\tau > w$ are preserved, but the scaling exponent within the segments is modified ($H \rightarrow H+1$). Additionally, a noisy environment was simulated by adding Gaussian white noise---different conditions were produced by changing the noise-to-signal ratio (NSR). The NSR is defined as the SD of the noise over the SD of the signal \cite{Olivares2016}. Finally, we integrate the series to estimate the normalized PE of the processes. Figure \ref{fig3} shows the estimated PE with $D=5$ (mean and SD) of twenty-one realizations as a function of the time scale $\tau$ for different noise levels and window sizes $w$. For the case of noise absence (NSR $= 0$, blue circles), we observe that PE converges slowly from a very low value to a {\it{plateau}} after the time scale equals the size of the window. In other words, for $\tau=1$, PE quantifies a strong persistence inside the windows due to the double integration. As $\tau$ increases, PE converges to a stable value (independent on $\tau$) quantifying a fBm with $H=5/6$ (black dots). As it has been proposed by Zunino {\it{et al.}}  \cite{Zunino2008}, the independence on the time scale evidences the presence of a self-similar stochastic process (fBm). On the other hand,  when dealing with a noisy environment, PE shows a minimum. Now, for $\tau=1$, the PE ``sees" the noise. The greater the noise level the greater the value of the normalized PE. As the time scale increases, the entropy quantifies the process (of length $w$) with a minimum, and finally it stabilizes to the value corresponding to a pure fBm with $H=5/6$. Similar results are obtained with $D=3$, 4 and 6. From this numerical analysis it can be concluded that PE is able to quantify the interplay between two different scaling laws in the same TS, even when the noise contamination is strong. 

This numerical analysis leads us to a better interpretation of the characterization depicted in Fig. \ref{fig2}. The analyzed sequences present a crossover between an integrated process for small time scales and a persistent fractal behavior for greater time scales. The independence of the PE on larger time scales evidences a self-similar stochastic series (fBm) on this temporal range \cite{Zunino2008}.   There exists electronic noise contamination, which is affected by the turbulent strength. As the intensity of the turbulence increases, the coordinates fluctuations are larger and, consequently, the related NSRs are smaller. This fact is clearly concluded from the comparison between experimental and simulation results (Figs. \ref{fig2} and \ref{fig3}). 

\begin{figure} [!ht] 
\centering
\includegraphics[scale=0.3]{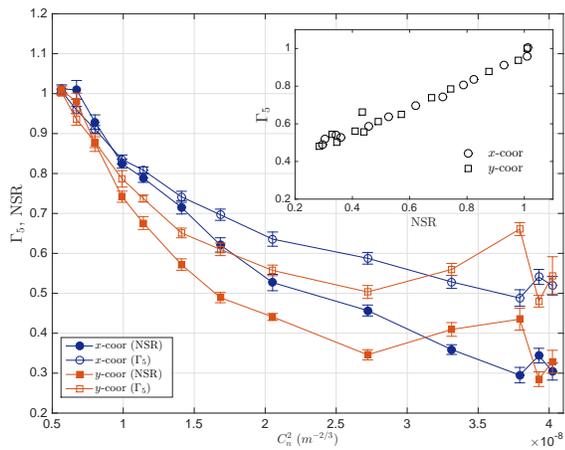}
\caption{Mean and SD (displayed as error bars) of the noise quantifier $\Gamma_5$ and NSR for the horizontal and vertical coordinates of the laser beam centroid as a function of the turbulence strength. In each case, both quantifiers are averaged over the twenty-one realizations. The inset plot shows $\Gamma_{5}$ versus NSR (only mean values are depicted for better visualization).}\label{fig4}
\end{figure}

With the purpose of quantifying the level of noise in the measurements from an entropic point of view, we define a noise quantifier,
\begin{equation}
	\Gamma_D =  \frac{(\mathcal{H}_D^1)_{\text{signal}}}{(\mathcal{H}_D^1)_{\text{noise}}},
\end{equation} \label{eq2}as the normalized PE associated with the signal at a given turbulent condition relative to the entropy of the background noise. Since we have concluded, from Fig. \ref{fig3}, that the entropy evaluated at $\tau=1$ detects the amount of noise, we evaluate both entropies in $\Gamma_{D}$ with this embedding delay. The background noise was considered as the reference measurements taken with both fans off and the heater disconnected, in order to characterize electronic noise and room turbulence effects only.  Figure \ref{fig4} depicts the estimated $\Gamma_{5}$ as a function of the turbulent strength. For the sake of comparison the NSR is also plotted. We observe that the noise quantification through $\Gamma_{5}$ is qualitatively equivalent to the results obtained by estimating the NSR. A similar qualitative characterization is obtained for other values of $D$. The inset plot compares both quantifiers. Regardless that $\Gamma$ depends on the parameter $D$, the linear dependence observed remains at least for $D=3$, 4 and 6. Clearly, as the turbulent intensity increases, the electronic noise contamination decreases.

\begin{figure} [!ht] 
\centering
\includegraphics[scale=0.3]{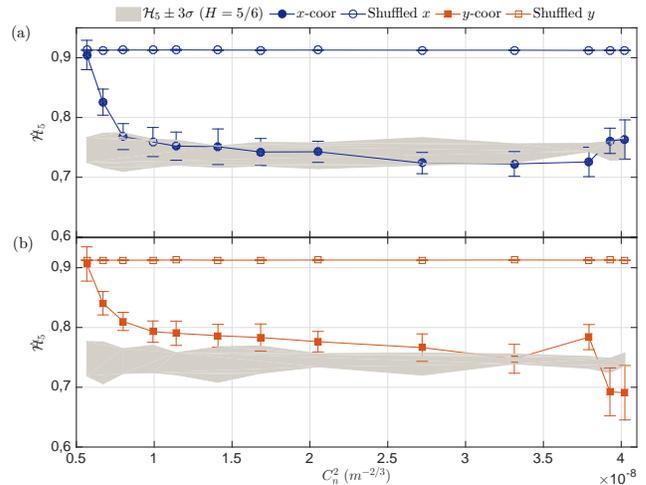}
\caption{Mean and standard deviation of $\hat{\mathcal{H}}_5$ from the interval $\tau \in [30, 50]$, as a function of the turbulent strength $C_n^2$ for (a) horizontal coordinate and (b) vertical coordinate. The gray region indicates $\mathcal{H}_5 \pm 3\sigma$ from a simulated fBm with $H=5/6$ contaminated with noise according to the NSR computed from the experiments (see Fig. \ref{fig4}). The entropy $\hat{\mathcal{H}}_5$ estimated from the shuffled data is also depicted. Qualitative similar results are found with $D=3$, 4 and 6.}\label{fig6}
\end{figure}

Now, we focus on the long-range correlations observed for large temporal scales. From Fig. \ref{fig2}, it can be concluded that after $\tau=30$, stability of $\mathcal{H}_5$ is reached. Since the constant value of the PE is a signature of a scale invariant stochastic process \cite{Zunino2008}, the fBm appears as a suitable model. We chose the interval $\tau \in [30, 50]$ to estimate a single value for the entropy, $\mathcal{\hat{H}}_5$. This result is depicted in Fig. \ref{fig6}. More precisely, mean and SD of the estimated PE over the 21 realizations, averaged over the interval $\tau \in [30, 50]$, are depicted. The gray region indicates a $3\sigma$ confidence interval ($\hat{\mathcal{H}}_5 \pm 3\sigma$) estimated from 21 independent realizations of a fBm with $H=5/6$ contaminated with noise according to the NSR computed from the experimental measurements. The similarity between the entropy estimated from horizontal and vertical coordinates sequences accounts for the isotropy of the turbulence within the laboratory chamber. As the turbulent intensity increases, PE saturates to the entropy value associated with a fBm with $H=5/6$ (gray region). This confirms a persistent fractal behavior of the laser beam wandering in laboratory-generated turbulence. As expected, the persistent behavior is absent when the entropy is estimated from the associated shuffled sequences, since all temporal structures are destroyed, obtaining a value for the PE associated for a random walk ($\mathcal{H}_{5} \sim 0.92$).

With the aim of studying the strong persistent observed (very low values of the PE) for small time scales, we consider the highest turbulent intensity, Fig. \ref{fig2}(e), because there exists the minimum noise contamination (no minimum observed in the normalized PE). For $\tau=1$ the averaged values ($\pm$ SD) of PE over the 21 realizations for the horizontal and vertical coordinates records are $ \langle \mathcal{H}_5 \rangle =0.48\pm 0.02$ and $ \langle \mathcal{H}_5 \rangle =0.5\pm 0.04$, respectively.  In addition, we have estimated the PE for one hundred independent  realizations of a fBm with $H \rightarrow 1$, and it is $ \langle \mathcal{H}_5  \rangle =0.52 \pm 0.02$. This is a lower bound of the PE for the fBm stochastic processes; no lower values are allowed if a fBm model is considered. This result indicates that the scaling behavior observed at small time scales could correspond to an integrated process. 

\begin{figure*} [!ht] 
\centering
\includegraphics[scale=0.45]{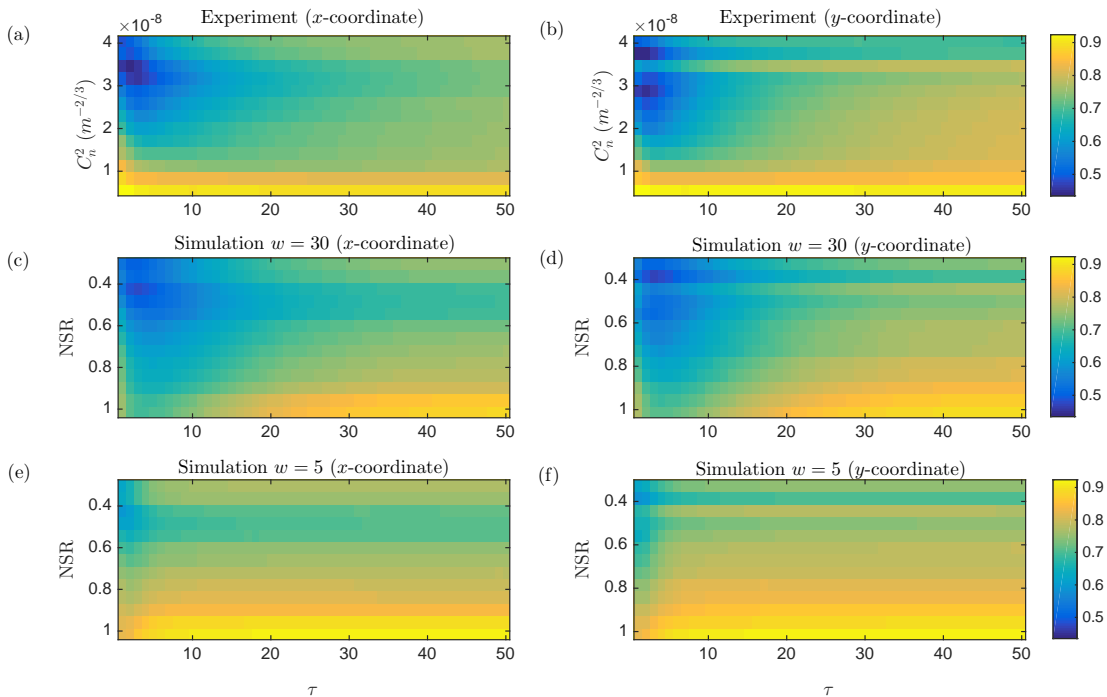}
\caption{Averaged normalized PE estimated with $D=5$ for the horizontal and vertical experimental records and for the numerical simulations. In the former case, the analysis is developed as a function of the embedding delay $\tau$ and the turbulence intensity while in the later case the NSR is implemented for quantifying the turbulence strength.}\label{fig5}
\end{figure*}

As a further analysis of the whole experimental data set, in Fig. \ref{fig5}(a)-(b) we depict the averaged PE, $\mathcal{H}_5$ over the twenty-one realizations, as a function of the time scale $\tau$, for all turbulence intensities considered in the experiment. In addition, Fig. \ref{fig5}(c)-(f) show the averaged entropy as a function of $\tau$ and the NSR, from a simulation of a noise contaminated fBms with a well-defined crossover at a time scale $\tau=30$ ((c)-(e)) and $\tau=5$ ((e)-(f)), between a process and long-range correlated fluctuations, as previously described. To simulate the fBms, we have used the averaged Hurst exponents estimated by using the traditional detrended fluctuation analysis (DFA) \cite{Kantelhardt2001}, from the twenty-one experimental measurements for each $C_n^2$. DFA methodology has been widely proved to be robust in the analysis of experimental data \cite{Zunino2015B}. For its implementation in MATLAB we recommend Ref. \cite{Ihlen2012}. A detrending polynomial of second order and 96 different scales $s \in [10,5000]$ equally spaced in the logarithmic scale were employed in the DFA implementation \cite{Zunino2015B}. Mean and SD of the exponents are outlined in Table \ref{tab1}. With a first sight of Fig. \ref{fig5}, it is well-established that PE exhibits a transition to a stable value as the time scale increases for almost all $C_n^2$. For the case of lowest turbulent intensity ($C_n^2= 5 \times 10^{-9} \,\, m^{-2/3}$) the detector is unable to resolve the fluctuation of the coordinates, consequently the entropy quantifies a fully uncorrelated electronic noise associated with the detector for all time scales with a value $\mathcal{H}_5 \sim 0.92$, which is the value corresponding to the entropy of a random walk. Since the noise level in the measurements is directly related to the turbulent conditions (see Fig. \ref{fig4}), we can qualitatively compare the evolution of the entropy as a function of $C_n^2$ (the one estimated from the measured sequences), with the evolution of the entropy with the NSR in the numerical simulation. From the simulations we conclude that for high turbulent intensities the crossover happens for a time scale $\tau \sim 30$. On the other hand, for the lowest intensities, the crossover is located at a shorter time scale ($\tau \sim 5$). It is worth remarking the good match between the experiments and the numerical simulation. 

\begin{table}[h!]
  \centering
      \caption{Mean (SD) of the Hurst exponents of the twenty-one realizations for both coordinate fluctuations and all turbulent intensities $C_n^2$.}
  \label{tab1}
  \begin{tabular}{ccc}
   \hline \toprule
    $C_n^2 \times 10^{-9}$ $(m^{-2/3})$ & $ \langle H_x \rangle $(SD) & $ \langle H_y \rangle $(SD)\\
    \hline\hline \midrule
    5.7 & 0.49(0.04) & 0.49(0.04)\\
    6.7 & 0.57(0.03) & 0.59(0.04)\\
    8.0 & 0.69(0.04) & 0.66(0.03)\\
    9.9 & 0.74(0.03) & 0.71(0.03)\\
    11.4 & 0.77(0.04) & 0.72(0.03)\\
    14.1 & 0.78(0.03) & 0.76(0.04)\\
    16.8 & 0.82(0.04) & 0.76(0.04)\\
    20.5 & 0.82(0.03) & 0.78(0.03)\\
    27.2 & 0.86(0.03) & 0.81(0.04)\\
    33.1 & 0.89(0.04) & 0.84(0.04)\\
    37.9 & 0.86(0.05) & 0.79(0.06)\\
    39.2 & 0.80(0.04) & 0.88(0.05)\\
    40.2 & 0.79(0.06) & 0.83(0.05)\\
	
     \hline \bottomrule
  \end{tabular}
\end{table}

\section{Conclusions}  \label{CC}

We have experimentally characterized the temporal structural diversity of the laser beam wandering in isotropic optical turbulence, through a multiscale permutation entropy analysis. In addition, we have performed a numerical analysis based on simulations of fBm stochastic processes, in order to interpret the experimental results. Our findings show that the multiscale permutation entropy characterizes all the relevant dynamical information contained in the laser beam coordinates fluctuations. The role played by the time scale $\tau$ is crucial to observe dynamical changes in the series. 

We have quantified the amount of noise contamination by defining a measure, namely $\Gamma_D$, as the ratio between the entropy of the signal over the entropy of the background noise for  $\tau=1$. We showed that $\Gamma_D$ is qualitatively equivalent to the NSR. From an entropic point of view, we demonstrated that the turbulent strength is related to the noise contamination. Thus, we were able to associate the NSR in the simulations with the intensity of the turbulence, $C_n^2$.

We have also demonstrated the presence of a crossover between two different scaling laws in the coordinates fluctuations. We observed an integrated fBm process, for small time scales, while for larger time scales we have identified a transition to a self-similar correlated time series. Actually, the laser beam wandering could be modeled by a persistent fractal behavior with a Hurst exponent $H=5/6$ for the stronger turbulences as the sampling time increases. The quantitative similarity between the entropy estimated for both, horizontal and vertical, coordinates fluctuations accounts for the isotropy of the turbulence within the laboratory chamber. Finally, we conclude that a multiscale estimation of PE is essential for uncovering all the dynamical information of the experimental measurements of a laser beam wandering.    

\section*{ACKNOWLEDGMENTS}
FO thanks support from Pontificia Universidad Cat\'olica de Valpara\'iso. DGP acknowledges financial support from Comisi\'on Nacional de Investigaci\'on Cient\'ifica y Tecnol\'ogica
(CONICYT), Chile (1140917, FONDECYT); Pontificia Universidad Cat\'olica de Valpara\'iso (PUCV), Chile (123.731/2014). LZ, DG and OAR gratefully acknowledge financial support from Consejo Nacional de Investigaciones Cient\'ificas y T\'ecnicas (CONICET), Argentina.

\end{document}